\documentclass[sort&compress]{elsarticle}

\usepackage{lineno,hyperref}
\usepackage{mathrsfs}
\usepackage{amssymb}
\usepackage{hyperref}
\usepackage{geometry}
\usepackage[section]{placeins}
\usepackage{graphicx}
\newcommand{\text}[1]{\mathrm{#1}}
\modulolinenumbers[5]

\journal{Journal of Magnetism and Magnetic Materials}

\begin{document}

\begin{frontmatter}

\title{Soliton motion induced along ferromagnetic skyrmion chains in chiral thin nanotracks}
\author{J. C. Bellizotti Souza$^1$}
\author{N. P. Vizarim$^{1, 2}$}
\author{C. J. O. Reichhardt$^3$}
\author{C. Reichhardt$^3$}
\author{P. A. Venegas$^4$}
\address{$^1$POSMAT - Programa de P\'os-Gradua\c{c}\~ao em Ci\^encia e Tecnologia de Materiais, Faculdade de Ci\^encias, Universidade Estadual Paulista - UNESP, Bauru, SP, CP 473, 17033-360, Brazil}
\address{$^2$Department of Physics, University of Antwerp, Groenenborgerlaan 171, B-2020 Antwerp, Belgium}
\address{$^3$Theoretical Division and Center for Nonlinear Studies, Los Alamos National Laboratory, Los Alamos, New Mexico 87545, USA}
\address{$^4$Departamento de F\'isica, Faculdade de Ci\^encias, Unesp-Universidade Estadual Paulista, CP 473, 17033-360 Bauru, SP, Brazil}

\begin{abstract}
  Using atomistic magnetic simulations we investigate the soliton motion along a pinned skyrmion chain containing an interstitial skyrmion. We find that the soliton can exhibit stable motion along the chain without a skyrmion Hall effect for an extended range of drives. Under a constant drive the solitons have a constant velocity. We also measure the skyrmion velocity-current curves and identify the signatures of different phases including a pinned phase, stable soliton motion, and quasi-free motion at higher drives where all of the skyrmions depin from the pinning centers and move along the rigid wall. In the quasi-free motion regime, the velocity is oscillatory due to the motion of the skyrmions over the pinning sites. For increasing pinning strength, the onset of soliton motion shifts to higher values of current density. We also find that for stronger pinning, the characteristic velocity-current shape is affected by the annihilation of single or multiple skyrmions in the drive interval over which the soliton motion occurs. Our results indicate that stable skyrmion soliton motion is possible and could be useful for technological applications.  
\end{abstract}


\end{frontmatter}

\section{Introduction}
There is great interest in finding solid state methods
that permit more rapid information transport.
Recent studies along
these lines involve the
use of topologically stable objects
as information carriers due to their stability,
low dimensionionality, and ability to be manipulated.
Notable examples used in both the fundamental and applied
research communities
include quantum Hall states,
chiral edge states of topological insulators, and massless Majorana modes.
\cite{DasSarma05,Moore10,Hasan10}.
In the context of magnetism, topological objects such as solitons, vortices, 
and monopoles are promising
information carrier
candidates due to their low-dimensional characteristics. 
Notably, the recent experimental observations of magnetic skyrmions in chiral
thin films or magnetic bulk crystals
\cite{Muhlbauer09,Yu10,Fert13}
have introduced a new class to the family of topological magnetic objects.
Due to their reduced size, easy current-driven motion, and topological
stability,
magnetic skyrmions are intensely studied
for the development of
new integrated circuits that can facilitate more reliable, 
compact, and energy-efficient information transport.

Magnetic skyrmions are topologically protected spin textures
\cite{Nagaosa13} that 
arise in magnetic materials and 
exhibit particle-like behavior under small driving forces 
\cite{Muhlbauer09,Yu10,Nagaosa13,Litzius17}.
Due to their smaller size and reduced energy cost, skyrmions are promising candidates for spintronics devices 
\cite{Nagaosa13,EverschorSitte18,Fert17}, such as magnetic logic gates
\cite{Luo18,Shu22,Zhang15},
transistors \cite{Zhang15c} and diodes
\cite{Souza22,Souza23,Feng22,Wang20c,Song21,Jung21,Shu22,Zhao20a}.
Many proposals for applications in spintronics devices require precise control of skyrmion motion 
\cite{Pfleiderer11,Wiesendanger16,Fert13,Zhang15,Luo18,Kang16a}.
Skyrmions share similarities with other overdamped particles, such as superconducting vortices, colloids, and electrons on Wigner crystals. These particles minimize their mutual interaction energy
by forming a triangular lattice,
and they can be driven by external forces while 
interacting with pinning sites in the material
\cite{Reichhardt17}. 
A key distinction between skyrmions and other overdamped 
particles is the presence of a strong non-dissipative Magnus force.
This force generates a velocity component perpendicular to
the net forces acting on the skyrmion, 
and the sign of this
velocity component depends on the
topological charge of the skyrmion
\cite{Nagaosa13,Iwasaki13,Litzius17,Jiang17,Lin13a,Lin13,Zeissler20}.
In clean samples, where there are no defects with which the skyrmions
can interact, the skyrmion motion occurs along an angle
$\theta_{sk}^{int}$, known as the skyrmion intrinsic Hall angle 
\cite{Nagaosa13,Iwasaki13,Litzius17,Jiang17,Lin13a,Lin13},
with respect to the externally applied driving force. 
Under specific conditions, the skyrmion Hall angle can be zero, 
depending on the phenomenological Gilbert damping parameter
and non-adiabatic spin transfer torque
\cite{Iwasaki13,Zhang17a}. 
The skyrmion Hall angle also depends on parameters such as the
skyrmion size, with experimental results showing skyrmion Hall angles
up to to $50^\circ$ 
\cite{Reichhardt17,Litzius17,Jiang17,Zeissler20,Brearton21}, but larger 
angles are also possible depending on material parameters
\cite{Nagaosa13,Reichhardt17}.
For many applications the skyrmion Hall angle can be a limiting factor since the skyrmions can travel to the edge of the sample and be annihilated there,
so there is great interest in identifying ways to move skyrmions under
conditions of zero skyrmion Hall angle.

To harness the potential of skyrmions for spintronic applications,
precise control of both individual and collective skyrmion motion
has been investigated intensively.
Various methods have been explored, such as periodic pinning 
\cite{Reichhardt15a,Reichhardt18,Feilhauer20,Vizarim21a,Vizarim20d,Vizarim20,Vizarim20a,Vizarim20c,Reichhardt10a},
ratchet effects \cite{Reichhardt15aa,Gobel21,Souza21,Chen19},
interface guided motion \cite{Vizarim21,Zhang22}, 
magnetic and temperature gradients
\cite{Yanes19,Zhang18,Everschor12,Kong13}, 
sample curvature \cite{CarvalhoSantos21,Korniienko20,Yershov22},
skyrmion-vortex coupling using a ferromagnetic-superconductor heterostructure
\cite{Menezes19}, 
skyrmion lattice compression
\cite{Zhang22a,Souza23a}, and laminar and turbulent flow
of skyrmions \cite{Zhang23}. 
In the case of skyrmions interacting with periodic substrates,
commensurability effects are
crucial for determining the dynamical behavior of
the skyrmions. 
Commensurability occurs when the
ratio of the number of skyrmions $N_{sk}$
to the number of pinning centers $N_p$ is an integer
or rational fraction.
The effects of commensurability
have been extensively studied in systems such as superconducting vortices 
\cite{Welp05,Reichhardt98,Harada96},
colloidal particles \cite{Mangold03}, Wigner crystals \cite{Rees12} and vortices
in Bose-Einstein condensates
\cite{Pu05,Tung06},
but exploration of commensurability in skyrmion systems has been limited
despite the expectation that new phenomena will arise due to the Magnus
force
\cite{OlsonReichhardt14,Reichhardt22b}.
When the system is close to but not at a commensurate state,
most of the sample
is ordered but there can be well defined
incommensurations in the form of solitons (kinks) or antisolitons
(antikinks) that act like quasiparticles with their own dynamics.
There can be an extended range of 
external drives where the solitons are depinned
while the the commensurate background remains pinned. 
The sliding dynamics of solitons on periodic substrates has been studied
for superconducting vortices \cite{Reichhardt98,Gutierrez09},
colloids \cite{Bohlein12,Vanossi12,Juniper15},
and frictional systems \cite{Benassi11}.

In a recent proposal, solitons moving through skyrmion
chains were used as information
carriers rather than the skyrmions themselves, and this
approach was explored
using a particle-based model \cite{Vizarim22}. 
The skyrmions are stabilized
in a sample with a periodic array of attractive defects, and by
adding or removing a skyrmion from one of the rows of the array,
a vacancy or interstitial soliton can be created.
Under an applied current, the soliton moves while
most of the skyrmions remain pinned. 
The main advantage of this technique
is that the soliton is stable and
can be set into motion with very low applied transport currents that are 
even smaller than the currents needed to depin
the bulk of the skyrmions. 
These findings pave the way for fast and energy-efficient information transport in potential skyrmion devices;
however, some important details remain to be explored, 
since the limitations of the particle model leave certain questions unanswered. 
For example,
it is not known if
the soliton can be stabilized when the skyrmions are not rigid but have
internal degrees of freedom, nor has the stability of the soliton
been tested in a dynamic context where skyrmion creation and/or annihilation
is possible.
Answering these and other questions requires a more detailed microscopic analysis.

In this work we use atomistic magnetic simulations to
investigate the dynamical behavior of a skyrmion chain slightly away from
commensuration, where an extra skyrmion is stabilized as an
interstitial.
In order to confine the skyrmions and guide their motion,
we introduce a
rigid wall above and below the line of pinning centers.
The interstitial skyrmion depins at lower driving forces than the
other skyrmions,
creating a soliton pulse that propagates through the sample.
In our work we provide a detailed description of the soliton motion and
analyze the different dynamical phases
that occur when the external driving force is varied.
By adjusting the pinning strength,
we determine the conditions under which the soliton can exist
as well as the annihilation conditions required to destroy the soliton 
motion.
In particular, we discover a reentrant pinning phase
that arises due to the annihilation of the interstitial skyrmion, 
which destroys the soliton motion and prevents further motion.
These new findings are only possible through a microscopic approach.
These results can be applied to the design of new skyrmionic devices
in which more efficient transport of information is achieved.

\section{Simulation}

\begin{figure}
\includegraphics[width=\textwidth]{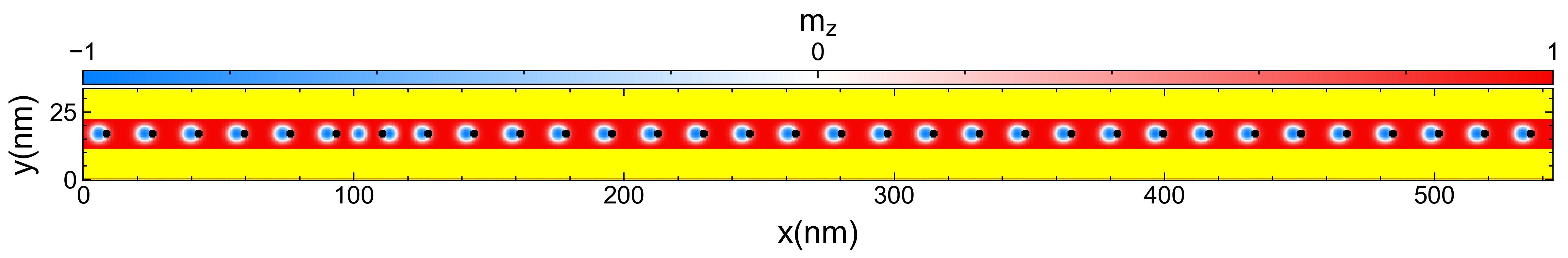}
\caption{Image of the starting spin configuration for all simulations.
Black dots are the pinning centers where
the easy-plane anisotropy $K<0$. Yellow represents rigid
magnetic walls with
$\mathbf{m}=\hat{\mathbf{z}}$.
Skyrmions with $m_z=-1$ are able to exist in the red region, where
the background magnetic field is $m_z = 1$.
We consider conditions under which there is one skyrmion per pinning center
along with one extra skyrmion
located in the space between two pinning centers
(shown at around $x=100$nm), giving a ratio of
$(32+1)/32=1.03$ skyrmions per pinning center.}
    \label{fig1}
\end{figure}

We consider a thin ferromagnetic nanotrack
of size
544 nm $\times 34$ nm
placed in the $x-y$ plane
that can support Neél skyrmions at $T=0$K under the influence of a
magnetic field applied
perpendicular to the film surface, as illustrated in Fig. \ref{fig1}.
The sample has periodic boundary conditions along the $x$ direction.
In this nanotrack, two rigid magnetic walls of thickness
$\Delta y = 14$ nm
are placed along $y=0$ nm and $y=34$ nm. 
Thus, the only place in which skyrmions
are able to reside is between the walls
in the red region where $m_z = 1$, as shown
in Fig.~\ref{fig1}. Embedded in the ferromagnetic nanotrack are
$N_p = 32$ equally spaced attractive pinning centers.
We investigate the dynamical behavior of a skyrmion chain
that is slightly away from commensuration
with $N_{sk}=33$ skyrmions. Note that due to the mismatch
$N_{sk}/N_p=1.03$ between skyrmions and pinning centers,
one skyrmion stabilizes in an interstitial position between pinning sites.

Our simulations are performed with the atomistic model
\cite{evans_atomistic_2018} 
for simulating individual atomic magnetic moments. The Hamiltonian
is \cite{iwasaki_current-induced_2013,Iwasaki13, seki_skyrmions_2016}: 

\begin{eqnarray}\label{Eq1}
\mathscr{H}=-\sum_{i, j\in N}J_{ij}\mathbf{m}_i\cdot\mathbf{m}_j
                -\sum_{i, j\in N}\mathbf{D}_{ij}\cdot\left(\mathbf{m}_i\times\mathbf{m}_j\right)
                -\sum_{i}\mu\mathbf{H}\cdot\mathbf{m}_i \nonumber\\
                -\sum_{i\notin P}K_0\left(\mathbf{m}_i\cdot\hat{\mathbf{z}}\right)^2
                -\sum_{i\in P}K\left(\mathbf{m}_i\cdot\hat{\mathbf{z}}\right)^2
\end{eqnarray}

The first term on the right hand side is the exchange interaction with
$N$ neighboring spins, and
the second term is the Dzyaloshinskii–Moriya interaction
for thin films.
The third term is the Zeeman interaction with an applied magnetic
field $\mathbf{H}$, where $\mu$ is
the magnitude of the atomic magnetic moments.
The last two terms are anisotropy interactions,
with the fourth term representing the sample anisotropy
and the fifth term the anisotropy produced by the set of $P$ locations
in which pinning centers reside.
%pinning centers anisotropy (set $P$).
The pinning center anisotropy
must be an easy-plane anisotropy in order to
attract the skyrmion border.
This is achieved
by selecting $K$ such that $K<0$ \cite{stosic_pinning_2017}.
The rigid walls confining the skyrmion chain are modeled as a fixed
atomic magnetic moment with $\mathbf{m}_{\in W}=\hat{\mathbf{z}}$
where $W$ is the set of locations composing the wall.

The time evolution
of the system is described by the LLG equation augmented with 
the adiabatic spin-transfer torque:
\cite{Iwasaki13, iwasaki_current-induced_2013, seki_skyrmions_2016, slonczewski_dynamics_1972, gilbert_phenomenological_2004}:
\begin{equation}\label{Eq2}
        \frac{d\mathbf{m}_i}{dt}=
        -\gamma\mathbf{m}_i\times\mathbf{H}^\text{eff}_i
        +\alpha \mathbf{m}_i\times\frac{d\mathbf{m}_i}{dt}
        +\frac{pa^3}{2e}\left(\mathbf{j}\cdot{\nabla}\right)\mathbf{m}_i
\end{equation}

Here $\gamma$ is the gyromagnetic ratio given by $\gamma=g\mu_B/\hbar$, $\alpha$
is the Gilbert damping parameter, $\mathbf{H}^\text{eff}_i=-\frac{1}{\mu}\frac{\partial \mathscr{H}}{\partial \mathbf{m}_i}$
is the effective magnetic field which encapsulates all interactions from the Hamiltonian, and the 
last term is the applied current where $p$ is the polarization,
$e$ is the electron charge, $a$ is the lattice
constant, and $\mathbf{j}$ is the applied current density.
The model for the current includes the assumption that the conduction electron
spins are parallel to the local magnetic moments $\mathbf{m}$ \cite{Iwasaki13, zang_dynamics_2011}.
This type of current is called the spin-transfer torque,
and its inclusion implies that the applied current is adiabatic.
Non-adiabatic terms
are not considered here since they do not affect the dynamical behavior
of rigid nanoscale skyrmions at small driving
forces \cite{Litzius17}, which is the regime we consider in this work.
We apply the current in the $+y$ direction, 
$\mathbf{j}=j\mathbf{\hat{y}}$, based on previous works detailing interactions
of skyrmions with magnetic walls \cite{Souza23,Souza22,Zhang22a}.

The number of skyrmions in the sample is measured
using topological charge calculations, and each skyrmion
has $Q=\pm1$ depending on the direction
of the applied magnetic field \cite{kim_quantifying_2020}.
The skyrmion velocities are calculated using the emergent electromagnetic fields \cite{seki_skyrmions_2016}: 
\begin{equation}
    E^\text{em}_i=\frac{\hbar}{e}\mathbf{m}\cdot(\partial_i\mathbf{m}\times\partial_t\mathbf{m})
    \;\;\;\;
    B^\text{em}_i=\frac{\hbar}{2e}\varepsilon_{ijk}\mathbf{m}\cdot(\partial_j\mathbf{m}\times\partial_k\mathbf{m})
\end{equation}
where $\varepsilon_{ijk}$ is the totally anti-symmetric tensor. The drift velocity of the skyrmions
is then computed according to $\mathbf{E}^\text{em}=-\mathbf{v}_d\times\mathbf{B}^\text{em}$ \cite{seki_skyrmions_2016, schulz_emergent_2012}.

Throughout the simulation we fix $\alpha=0.3$, $p=1.0$ and $a=0.5$nm. For
the material parameters we choose $J=1$ meV, $D=0.18J$ and $K_0=0.02J$.
The applied field is $\mu\mathbf{H}=0.5(D^2/J)\mathbf{\hat{z}}$, corresponding
to the skyrmion phase \cite{Iwasaki13, seki_skyrmions_2016}.
This field creates skyrmions with $Q=-1$ topological
charge.

In all simulations we start our system with a spin configuration, illustrated
in Fig \ref{fig1}, obtained by inducing the formation of 33 skyrmions and 
then relaxing the LLG equation in Eq. \ref{Eq2} with $\mathbf{j}=0$ for
several time steps until reaching 
a steady configuration.
After the sample has been initialized,
we apply a spin current $\mathbf{j}\neq0$ and begin
measuring the dynamics of the system. We integrate Eq.~\ref{Eq2} using a
fourth order Runge-Kutta method.
Simulation time units are normalized to
$t=(\hbar/J)\tau$ and current density to 
$\mathbf{j}=(2eJ/a^2\hbar)\mathbf{j}'$.
The values of time and current density reported here have been converted
back into dimensional physical units.

\section{Soliton Motion}

\begin{figure}
\includegraphics[width=\textwidth]{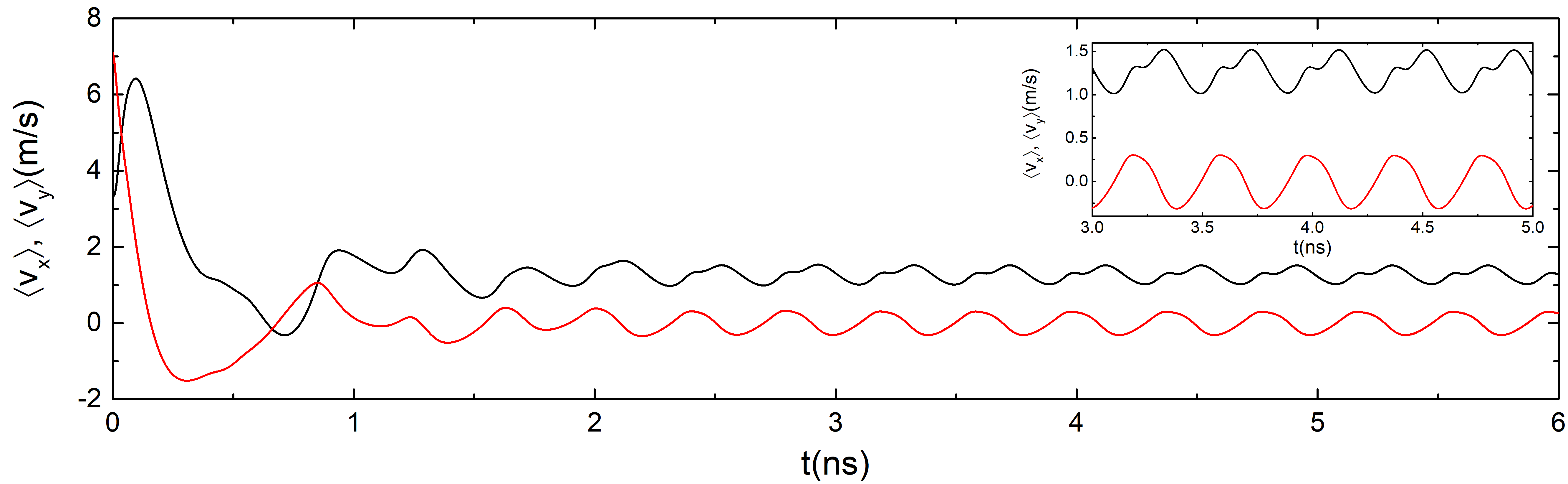}
\caption{Skyrmion average velocity signals
$\left\langle v_x\right\rangle$ (black) and
$\left\langle v_y\right\rangle$ (red)
for a system with $K=-0.02J$ and
$j=2.04\times10^{10}\text{A/m}^2$.
Here the plot is truncated
%    For clarity reasons the plot is truncated
after $t=6$ ns, but the pattern established in both
$\left\langle v_x\right\rangle$ and $\left\langle v_y\right\rangle$
repeats until the end of the simulation at
$t=59.2$ ns.}
\label{fig2}
\end{figure}

We first consider the system shown in Fig. \ref{fig1} using $K=-0.02J$ and $j=2.04\times10^{10}\text{A/m}^2$.
In this case, the transport current is weak enough so that skyrmion
annihilation never occurs.
In Fig.~\ref{fig2} the skyrmion average velocity signals
$\left\langle v_x\right\rangle$ and
$\left\langle v_y\right\rangle$ are plotted as a function of time, $t$.
Initially, for $t<1$ ns, there is a transient motion while
the skyrmions adjust themselves to the presence of the
applied transport current
combined with the rigid walls and pinning centers. 
Such transient motion occurs for low values of applied transport current
and is crucial for establishing a dynamical stabilization that
prevents skyrmion annihilation.
For stronger currents, 
the transient stabilization process is more abrupt and annihilation may occur.
Once $t>1$ ns, 
the skyrmions always exhibit an average velocity
$\left\langle v_x\right\rangle > 0$ indicating that
collective skyrmion motion
is occurring.
For the range $1 < t< 2.5$ ns, the velocities are not completely periodic since the transient stabilization of the motion is not yet
complete; however,
for $t>3$ ns, the skyrmion velocities exhibit a smooth periodic behavior as
shown in the inset of Fig.~\ref{fig2}.
Note that while $\left\langle v_x\right\rangle$ oscillates around $\left\langle v_x\right\rangle \approx 1.25$ m/s, $\left\langle v_y\right\rangle$
oscillates around zero, meaning that there is a net motion along $x$
but only bounded periodic excursions along $y$.
The steady state
$\left\langle v_x\right\rangle$ curve consists of repetitions of
two peaks and one valley.
This motion is associated with a soliton in
the skyrmion chain, where the interstitial skyrmion
pushes its neighboring pinned skyrmion out of 
a pinning center and takes up residence in the pinning center.
The previously pinned skyrmion becomes the new interstitial.

\begin{figure}
\centering
\includegraphics[width=0.5\textwidth]{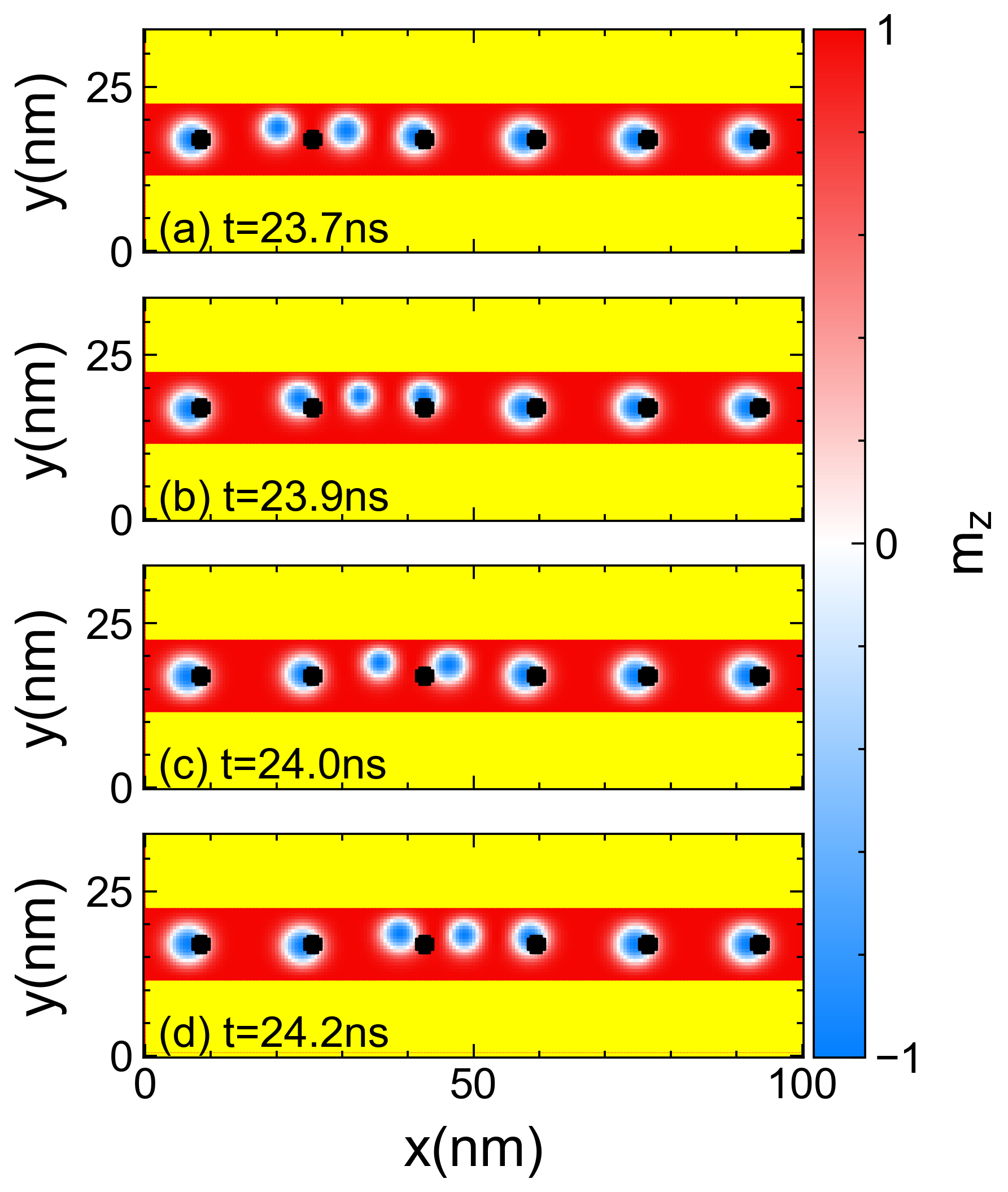}
\caption{Snapshots of the
spin configuration time evolution for a sample with
$K=-0.02J$ and $j=2.04\times10^{10}\text{A/m}^2$.
(a) $t=23.7$ ns.
(b) $t=23.9$ ns.
(c) $t=24.0$ ns.
(d) $t=24.2$ ns.
The soliton moves from its location between the second and third pinning
sites from the left to a new location between the third and fourth
pinning sites from the left.
%    From (a) to (d) the soliton motion around two pinning centers, with
%    $t=23.7$ns, $t=23.9$ns, $t=24.0$ns and $t=24.2$ns respectively.
An animation of this motion is available
in the supplementary material ``mov-fig3.mp4''.}
 \label{fig3}
\end{figure}

In Fig.~\ref{fig3} we illustrate the soliton motion from one interstitial
site to another 
using snapshots of a portion of the sample taken at consecutive values of $t$.
At $t=23.7$ ns in Fig.~\ref{fig3}(a),
the interstitial skyrmion
causes a previously pinned skyrmion to depin
due to the repulsive skyrmion-skyrmion interaction.
In Fig.~\ref{fig3}(b) at $t=23.9$ ns,
the interstitial skyrmion becomes pinned,
and the previously pinned skyrmion becomes the new interstitial
skyrmion.
In Fig.~\ref{fig3}(c) and (d)
the same process described in
Fig.~\ref{fig3}(a) and (b) repeats for the next pair of
skyrmions to the right,
indicating the cascading process of depinning 
and pinning of skyrmions.
The two peaks
in $\left\langle v_x\right\rangle$ shown in the inset of
Fig.~\ref{fig2}
correspond to the depinning
of a skyrmion from a pinning center
(weaker peak) and the velocity boost
caused by the interaction with
the rigid wall above (stronger peak).
The valley
in $\left\langle v_x\right\rangle$
occurs when
the moving interstitial skyrmion approaches
its pinned neighbor skyrmion and
slows down while pushing it until
the pinned skyrmion begins to move.
This process occurs as a repeating sequence,
as demonstrated in Vizarim \textit{et al.}
using a particle-based model \cite{Vizarim22}.
Here we show that this soliton motion remains robust even when
using a more detailed
micromagnetic model.

For the system illustrated in Figs. \ref{fig2} and \ref{fig3},
it takes roughly 0.5 ns for the soliton to translate from one interstitial
site to the next.
%the soliton displacement from one pinning center to another has a duration
%of roughly $0.5$ ns, according to Fig \ref{fig3}.
The distance between pinning sites
is $17$ nm. Thus, the soliton velocity is
$\left\langle v_s\right\rangle \approx 34$ m/s along
the $x$ direction.
Comparing this value to the average skyrmion velocity
shown in Fig.~\ref{fig2},
we find that the soliton moves much faster than the
individual skyrmions.
This indicates that the soliton could be employed as
a very fast information carrier using low applied transport currents.
%The supplementary material "mov-fig3.mp4" presents
%an animation for the soliton motion present on Fig \ref{fig3}.
%A visualization for the soliton motion focusing on skyrmion
%velocity are plotted on Fig \ref{fig4}.

\begin{figure}
\centering
\includegraphics[width=3.5in]{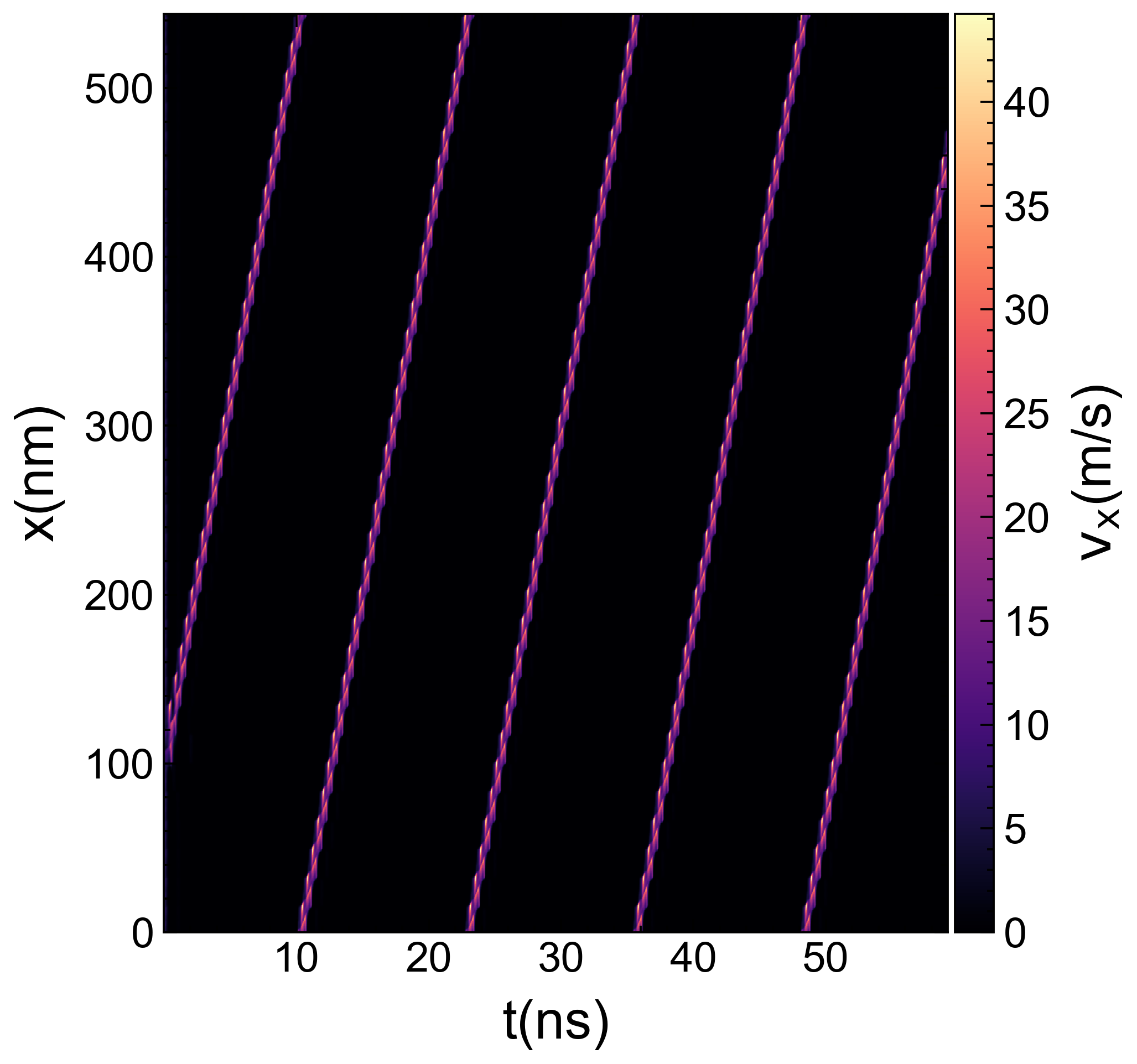}
\caption{Heat map of skyrmion velocity
$v_x$ as a function of $x$ versus time $t$ for the
system in Fig~\ref{fig3}. Brighter colors represent faster skyrmions.
The periodic discontinuities of the lines
are produced when the skyrmion passes through the periodic
boundary conditions in the $x$ direction.}
\label{fig4}
\end{figure}

Fig \ref{fig4} shows a heat map
of the skyrmion velocity $v_x$
plotted as a function of $x$ position versus time $t$.
This plot enables us to visualize the soliton motion
through the sample and record its velocity.
The periodic boundary conditions cause a
discontinuity of the soliton motion
each time the soliton exits the
right side of the simulation box and reenters on the opposite side.
As can be seen, all skyrmions except the ones supporting
the soliton pulse are pinned and therefore have $v_x=0$ m/s, whereas
the skyrmions under the soliton
have $v_x\neq 0$ m/s.
The area of finite $v_x$
%This effect when viewed on a heatmap plot create regions
%with $v_x\neq0$m/s, showing the
indicates the trajectory of the perturbation over time.
An interesting characteristic of the soliton motion
is its linear behavior. As shown in Fig \ref{fig4},
the soliton perturbation moves
across the sample with a constant velocity over time.
It is also possible to estimate the soliton velocity
from this plot since the soliton traverses the entire
simulation box of $544$ nm in roughly $14$ ns,
giving $\left\langle v_s\right\rangle \approx 38$ m/s.

\section{Magnitude of Transport Current}

\begin{figure}
\centering
\includegraphics[width=0.7\textwidth]{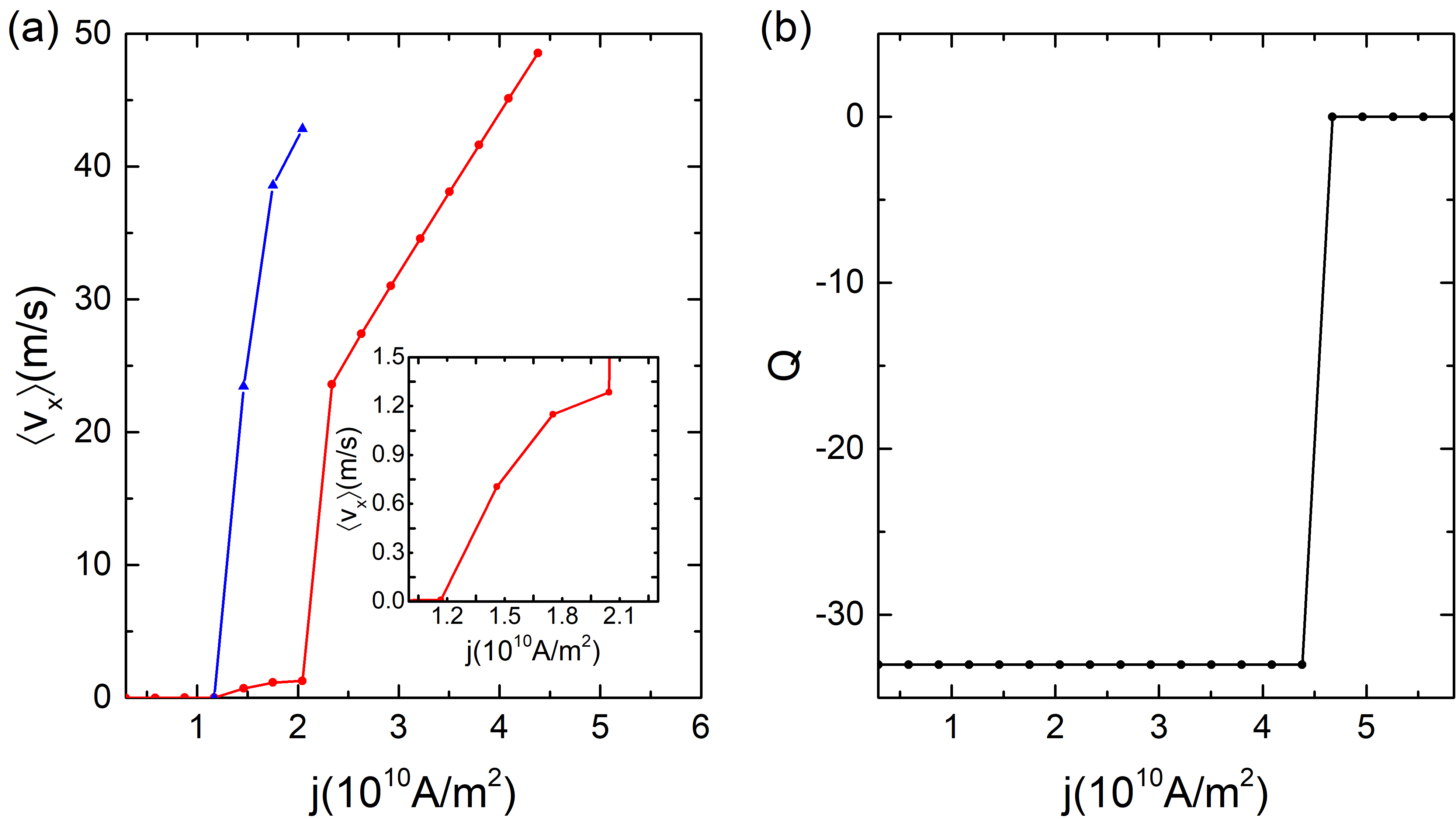}
\caption{ (a) The average velocity
$\left\langle v_x\right\rangle$ vs applied transport density $j$
for the skyrmions (red) and the soliton (blue).
(b)
The corresponding $Q$ vs $j$.
%The blue curve is the average velocity of the soliton pulse, not the skyrmion themselves.
The inset of (a) shows a
blowup of the average skyrmion velocity $\left\langle v_x\right\rangle$
vs $j$
focusing on the soliton regime in the interval of
$1.1\times10^{10}\leq j\leq 2.3\times10^{10}\text{A/m}^2$.
In (b) the topological charge changes from $Q=-33$ to $Q=0$ in a sharp step
around $j=4.4\times10^{10}\text{A/m}^2$, well above the value of $j$
for which all of the skyrmions have depinned.}
\label{fig5}
\end{figure}

In the last section we demonstrated that the soliton motion can be stabilized in the ferromagnetic nanotrack shown in Fig. \ref{fig1} for a specific
value of transport current. Now we investigate the stability
of the soliton motion as the transport current is
modified for fixed $K=-0.02J$.
We vary the current density magnitude
over the interval $2.9\times10^{9}\text{A/m}^2\leq j\leq 5.8\times10^{10}\text{A/m}^2$.
%as shown in Fig.~\ref{fig5}.
Figure~\ref{fig5}(a) shows the velocity signal $\left\langle v_x\right\rangle$
as a function of the applied transport current magnitude $j$. The inset in Fig \ref{fig5}(a)
is a blowup of
the main panel over
the interval $1.1\times10^{10}\leq j\leq 2.3\times10^{10}\text{A/m}^2$.
Three dynamic phases appear.
For low currents of $j<1.2\times10^{10}\text{A/m}^2$,
there is a pinned phase where
all of the skyrmions in the sample are pinned and the interstitial skyrmion
remains trapped by the caging potential
created by its neighboring pinned skyrmions.
In the interval
$1.2\times10^{10}\leq j\leq 2.1\times10^{10}\text{A/m}^2$, highlighted
in the inset of Fig.~\ref{fig5}(a), we find
a soliton motion phase,
where the soliton perturbation moves along the
skyrmion chain at a high velocity.
For the range $2.1\times10^{10}<j\leq5.8\times10^{10}\text{A/m}^2$,
we observe
quasi-free skyrmion motion where all
of the skyrmions depin and flow through the sample
at increased velocity.
Interestingly, the
average soliton velocity increases monotonically with $j$ until the
soliton is destroyed,
and the average soliton velocity is
always higher than the
average skyrmion velocity.
The shape of the
skyrmion velocity-current curve is consistent with previous
work on soliton motion in skyrmion chains
performed with a particle model \cite{Vizarim22}.
Note that here, we use a microscopic model
rather than a phenomenological model for
the interaction between skyrmions and pinning centers.
According to the
pinning center potentials described in Ref.~\cite{stosic_pinning_2017}, 
pinning centers created using variations in
anisotropy are attractive
to skyrmions off-center,
meaning that they attract the skyrmion domain wall
but repel the skyrmion core.
As a result, the skyrmions prefer to sit off-center in the pinning site.
On the other hand, the
phenomenological pinning centers used in
Ref.~\cite{Vizarim22} were modeled as parabolic traps that
attract
the skyrmion center, 
allowing the skyrmions
to become pinned precisely at the center of pins.
This small but significant difference in the skyrmion-pinning
interaction may result in different behaviors.

In Fig.~\ref{fig5}(b) we plot the stable topological charge $Q$,
obtained after the initial transient annihilation process is
complete, as a function
of applied current density $j$. 
%The stable topological charge
%is given by the value of $Q$ after the
%initial transient annihilation process is complete. 
As can be seen in Fig.~\ref{fig1},
there are 33 skyrmions in the sample, giving $Q=-33$. The value is negative
due to the direction of the in-plane magnetization of the skyrmion core.
Combining this information with
the curve in Fig.~\ref{fig5}(b), we can 
conclude that no skyrmions are annihilated for
$j\leq 4.7\times10^{10}\text{A/m}^2$.
Thus all three of the phases described above overlap with
current density regions where the number of skyrmions is stable as
a function of time.
%For the system shown in Fig. \ref{fig5}, there is no annihilation
%during the three phases. However,
Once the annihilation process begins,
however, it is irreversible since all of the skyrmions 
in the sample will annihilate if given sufficient time.
%This is highly depended on $K$,
%by interactions with pinning centers during quasi-free skyrmion motion.
%For $K=-0.02J$ the interaction is weak enough such that when
%nnihilation process starts all skyrmions are annihilated.

\begin{figure}
\includegraphics[width=\textwidth]{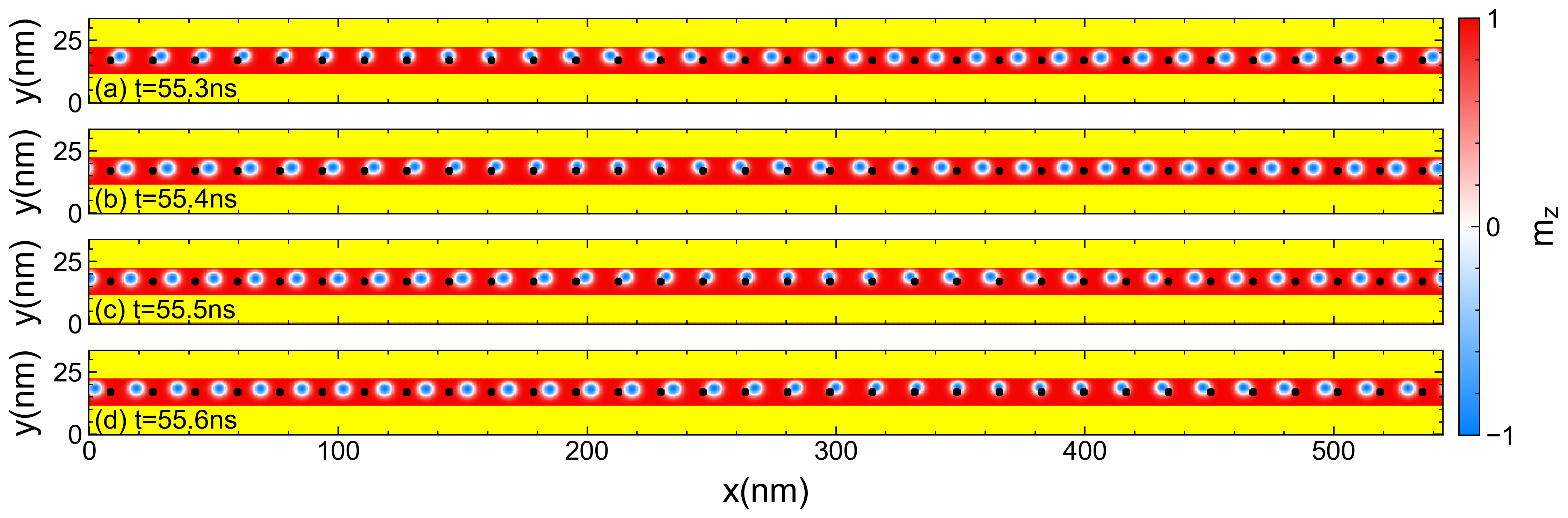}
\includegraphics[width=0.9\textwidth]{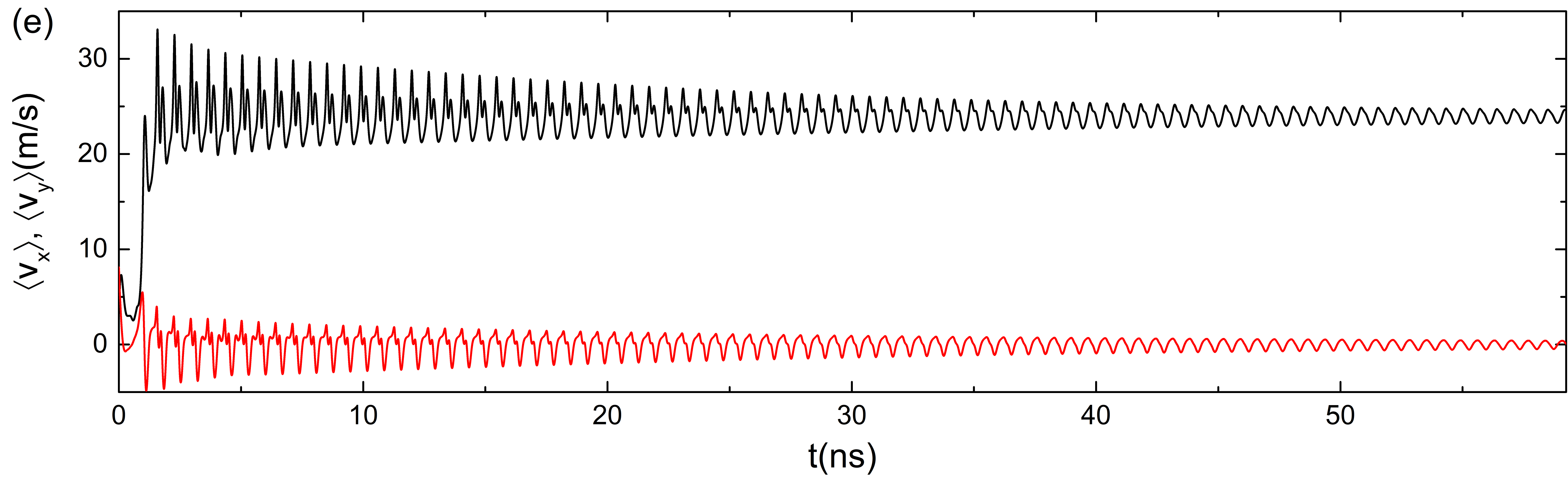}
\caption{(a-d) Snapshots of the time evolution of the spin
configuration for a system with $K=-0.02J$ and $j=2.3\times10^{10}\text{A/m}^2$,
where all of the skyrmions
have depinned from the pinning centers.
(a) $t=55.3$ ns.
(b) $t=55.4$ ns.
(c) $t=55.5$ ns.
(d) $t=55.6$ ns.
An
animation showing the complete time evolution is available in the
Supplemental Material as ``mov-fig6.mp4''.
(e) Velocity signals $\left\langle v_x\right\rangle$ (black) and
$\left\langle v_y\right\rangle$ (red) vs time $t$ for the same system.}
\label{fig6}
\end{figure}

The interaction with pinning centers during the quasi-free skyrmion motion
is illustrated in detail in Fig~\ref{fig6} for
the system from
%In Fig \ref{fig6} it is presented the details of the quasi-free skyrmion motion for the system of
Fig.~\ref{fig5} at
$K=-0.02J$ and $j=2.3\times10^{10}\text{A/m}^2$.
The time evolution of
the spin configuration in Fig.~\ref{fig6}(a-d)
shows that the transport current is large enough to depin all of
the skyrmions from the pinning centers at the same time.
The skyrmions still feel the influence of the pinning sites, however,
creating a perturbation that propagates through 
the skyrmion chain.
A portion of the skyrmions are closer to the
pinning sites and have reduced velocities.
Meanwhile, other skyrmions are further from the pinning sites
and flow faster.
The combination of these two motions and the interchange of skyrmions
between them creates a wave that propagates through the nanotrack.
This motion 
has a much higher velocity
than what we observe for the interstitial soliton.
%when compared to the interstitial soliton motion case.
%NICOLAS COMMENT: IS IT A TYPE OF SOLITON TOO? A TYPE OF MULTISKYRMION SOLITON? IS IT INTERESTING TO EXPLORE FURTHER?
%Cynthia says: the transport current is higher here so it is not surprising
%that the motion would be faster. Solitons can be partially delocalized so
%this could probably be modeled as an extended soliton object. In fact it
%would not surprise me if your soliton got "longer" (contained more particles)
%as j increases.
Fig.~\ref{fig6}(e) shows the time dependent velocities
$\left\langle v_x\right\rangle$ and $\left\langle v_y\right\rangle$
for this system.
%velocity signal for this
%system.
The initial motion is very chaotic and generates many peaks and valleys
in $\left\langle v_x\right\rangle$ and $\left\langle v_y\right\rangle$.
As time passes the system settles into periodic motion, and
for $t>45$ ns
%starts to behave periodically.
both $\left\langle v_x\right\rangle$ and $\left\langle v_y\right\rangle$
oscillate around stable values of
$\left\langle v_x\right\rangle \approx 24$ m/s and 
$\left\langle v_y\right\rangle \approx 0$ m/s. 
%The animation for this system is present
%in the supplementary material as "mov-fig6.mp4".

\section{Pinning Center Strength}

\begin{figure}
\centering
\includegraphics[width=0.9\textwidth]{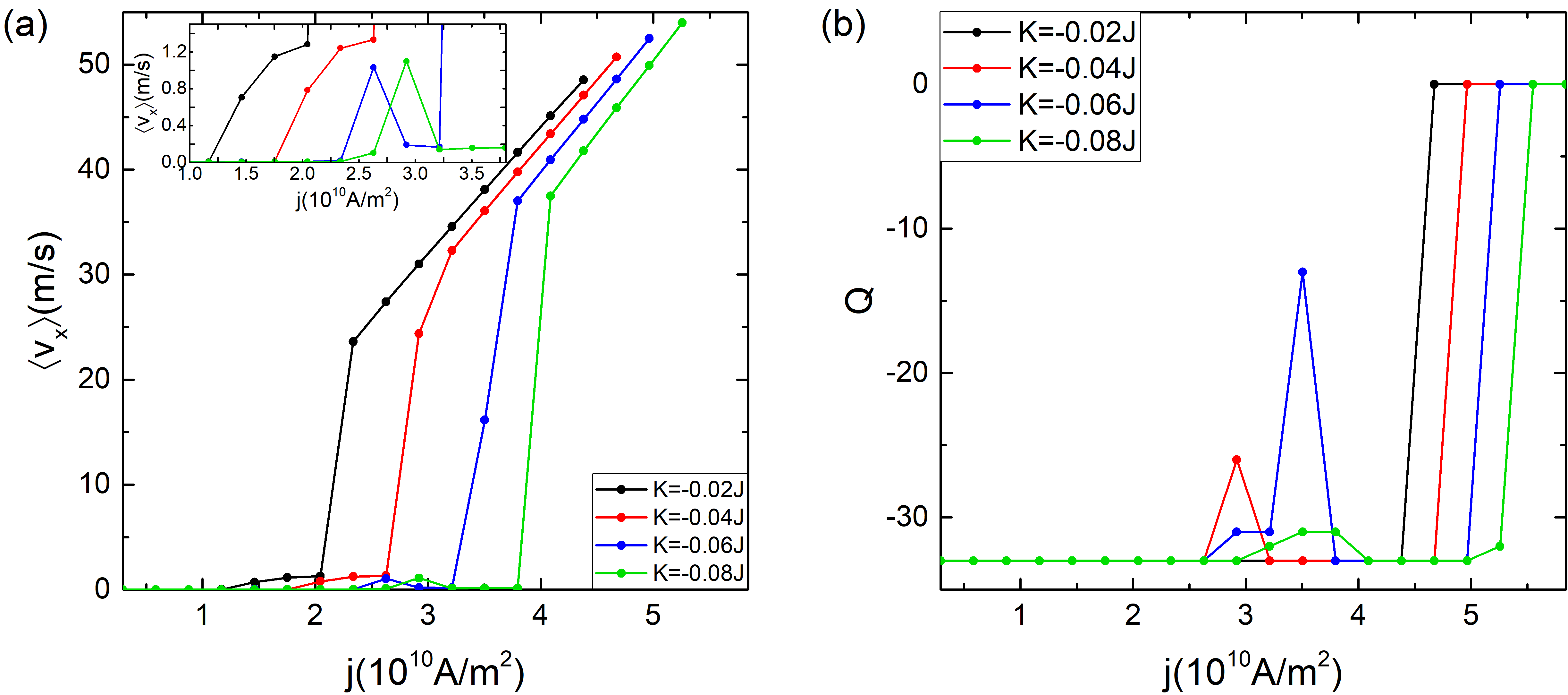}
\caption{Plots of (a) $\left\langle v_x\right\rangle$ vs $j$ and
(b) $Q$ vs $j$ for
four values of pinning strength,
$K=-0.02J$ (black), $-0.04J$ (red), $-0.06J$ (blue),
and $-0.08J$ (green).
The inset of panel (a) shows a blow up of the soliton motion regime.
As the
pinning strength increases, the onset of soliton motion
shifts to higher values of $j$.
For very strong pinning, $|K|\geq0.06J$, the
soliton motion disappears and is replaced
by a reentrant pinning phase due
to skyrmion annihilation.
The annihilation process is visible in
panel (b) where,
due to the annihilation of two
skyrmions or one skyrmion for $K=-0.06J$ and $K=-0.08J$, the system becomes
pinned again until stronger currents force all of the skyrmions
to move.
An animation showing the reentrant pinning phase is available
in the Supplemental Material as
``mov-fig7.mp4''.}
\label{fig7}
\end{figure}

In this section we investigate how the anisotropy strength of the
pinning sites affects the dynamics. We choose
values of $K$ from the range $-0.08J\leq K\leq-0.02J$. 
For the easy-plane anisotropy we require $K<0$,
and larger values of $|K|$ enhance the easy-plane anisotropy.
That is, larger values of $|K|$
produce a stronger attractive interaction
between a skyrmion and a pinning site.
%The velocity and stable topological charge plots as a function of $j$
%are present in Fig \ref{fig7} for some values of $K$.
Figure~\ref{fig7}(a) shows
the velocity $\left\langle v_x\right\rangle$ versus
applied current density $j$
for specific values of $K$.
The inset of Fig.~\ref{fig7}(a) shows a blow up
of the soliton motion
regime. As $|K|$ increases, the onset of soliton motion
shifts to higher values of $j$.
Such behavior is expected since stronger skyrmion-pinning
interactions make it necessary to
apply a higher transport current in order to depin the skyrmions.
The depinning currents are $j_c=1.2\times10^{10}\text{A/m}^2,
1.8\times10^{10}\text{A/m}^2, 2.3\times10^{10}\text{A/m}^2$ and $2.6\times10^{10}\text{A/m}^2$ for
$K=-0.02J, -0.04J, -0.06J$ and $-0.08J$, respectively.
Note that there is an intermediate point
of increased $\left\langle v_x\right\rangle$ for the $K=-0.08J$ curve,
visible in the inset of Fig.~\ref{fig7}(a).
This point 
does not indicate the existence of soliton motion.
Instead, there is only a transient motion
that ceases after $25$ ns.
As discussed above, when $|K|$ increases, the pinning centers become
more attractive for the skyrmion domain wall,
shifting the depinning transition to a larger applied current 
density.

The shape of the
skyrmion velocity-current curve
shown in Fig.~\ref{fig5} is preserved only for $|K|\leq 0.04J$.
In this regime,
the velocity-current curve exhibits the three distinct dynamic phases
discussed in the previous sections:
(i) the pinned phase,
(ii) soliton motion, and (iii) quasi-free skyrmion motion.
In contrast, for $|K|> 0.04J$ the velocity exhibits a spike
followed by a strong drop.
The depinning transition, velocity spike, and
velocity drop are shifted as the transport current density increases.
In this regime, four dynamic phases are present:
(i) the pinned phase, (ii) soliton motion,
(iii) a reentrant pinned phase,
and (iv) quasi-free skyrmion motion.
Phases (i), (ii), and (iv) are
the same as those already described in
the previous section, but the reentrant pinning phase is different.
During the reentrant pinning phase,
skyrmions are initially flowing
following the previous soliton motion,
but then the interstitial skyrmion, and in some cases more skyrmions depending on $|K|$,
is annihilated inside the sample due to the stronger pinning
produced by larger $|K|$. 
As a result of the annihilation, the soliton
is destroyed and the remaining skyrmions
become trapped at the pinning sites,
leading to the emergence of a pinned state.
Note that $\left\langle v_x\right\rangle \neq 0$ 
since the skyrmion is annihilated in this process,
so that $\left\langle v_x\right\rangle \approx 0$ but we do not have
$\left\langle v_x\right\rangle = 0$.

In Fig.~\ref{fig7}(b) we plot the topological charge
as a function of applied current density. For
$K=-0.02J$ there is only a single transition
in $Q$. The number of skyrmions is constant
until $j=4.4\times10^{10}\text{A/m}^2$,
above which all skyrmions are annihilated.
For $K=-0.04J$, at $j=2.9\times10^{10}\text{A/m}^2$
seven skyrmions are annihilated, 
corresponding to the point between the soliton motion
and the quasi-free motion shown in Fig.~\ref{fig7}(a).
It is important to note that for each value of $j$,
the initial skyrmion configuration is the
same as that illustrated in Fig.~\ref{fig1}, meaning that the
applied current is not swept up from zero.
The goal is to simulate a sample with an interstitial skyrmion
at different values of $j$ to see how it behaves.
When $K=-0.04J$,
for $2.9\times10^{10}<j<5\times10^{10}\text{A/m}^2$
no additional annihilation is observed,
while for $j>5\times10^{10}\text{A/m}^2$ all
of the skyrmions annihilate.
For $K=-0.06J$ there are two current densities at which
two skyrmions are annihilated ($j=2.9\times10^{10}\text{A/m}^2$ and
$j=3.2\times10^{10}\text{A/m}^2$). The annihilation of these two skyrmions
terminates the soliton motion
and causes $\left\langle v_x\right\rangle$ to drop to
$\left\langle v_x\right\rangle \approx 0$. 
For $j=3.5\times10^{10}\text{A/m}^2$, 20 skyrmions are annihilated.
This is the $j$ value just before the quasi-free skyrmion motion, indicating that the reentrant pinning phase is a transient phase between
the soliton and the quasi-free regime.
For larger current density values
$j\geq 5.3\times10^{10}\text{A/m}^2$, all of the skyrmions
are annihilated for $K=-0.06J$.
At $K=-0.08J$, there is a region where few skyrmions are 
annihilated,
annihilation of a single skyrmion for $j=3.2\times10^{10}\text{A/m}^2$, and
annihilation of two skyrmions
for $j=3.5\times10^{10}\text{A/m}^2$ and $3.8\times10^{10}\text{A/m}^2$.
The skyrmion annihilation
destroys the soliton motion and generates the reentrant pinning phase.
For $j\geq 5.6\times10^{10}\text{A/m}^2$ all of the skyrmions
in the sample are annihilated.
The onset of complete skyrmion annihilation shifts to
higher values of $j$ with increasing
$|K|$. When the interaction between skyrmions and pinning centers
is stronger, the interstitial skyrmion requires more energy
to push a pinned skyrmion from its pinning site.
As a consequence, the interstitial skyrmion interacts with the rigid wall
of the pinned skyrmion for a longer period of time,
increasing the chances of annihilation.

\section{Summary}
Using an atomistic model for atomic magnetic moments,
we simulated the dynamical behavior
of a skyrmion chain in a nanotrack embedded with pinning centers
and encased by rigid magnetic walls to guide the
skyrmion motion along the $x$ direction.
The system is slightly off the commensuration ratio,
resulting in the appearance of an
interstitial skyrmion that creates a high mobility 
soliton in the lattice.
As we apply a transport current density,
the interstitial skyrmion pushes its neighboring pinned skyrmion
into an interstitial position, and the previously
interstitial skyrmion becomes pinned by the
now vacant pinning site. This process occurs in a repeating sequence.
The soliton motion is the result of
a combination of interactions between the interstitial skyrmion,
rigid walls, pinning centers, and pinned skyrmions. 

We identified two peaks and a valley in the
skyrmion velocity signal as a function
of time.
The first, weaker peak
corresponds to the moment when a pinned skyrmion
depins from the pinning center due to its interaction with the
arriving interstitial skyrmion. 
The second, stronger peak arises
from a velocity boost due to the interaction of the skyrmion
with a rigid wall. 
The valley represents the point at which the interstitial
skyrmion
slows down because it has begun
to interact with the next pinned skyrmion in the chain.
The soliton motion repeats indefinitely in a periodic manner,
with the period controlled by the pinning density, 
illustrating the stability of the soliton motion.
Using a velocity heat map plotted as a function of position and time,
we observe that the soliton motion is linear in time with a constant velocity. 
This precise and controllable motion can be of value for technological applications where precise control of skyrmion motion is essential.

By varying the applied current density
we obtain a range of $j$ values for which the soliton motion
is stable. For $K=-0.02J$, soliton motion exists over the range
$1.2\times10^{10}\text{A/m}^2\leq j\leq2.1\times10^{10}\text{A/m}^2$,
and is accompanied by a nonlinear velocity-current signature
consistent with
what was found in previous particle-based studies of
soliton motion in skyrmion chains. 
We also show that the soliton moves much more rapidly
than the individual skyrmions, and thus
solitons
have promise for use as fast and energy-efficient information carriers, since they can 
be set into motion using very low external current densities.
For $j<1.2\times10^{10}\text{A/m}^2$ the system remains pinned since the
external current density is too low to overcome the pinning force.
When $j>2.1\times10^{10}\text{A/m}^2$, the soliton motion
is destroyed and
the velocity-current curve increases linearly with $j$
since all the skyrmions
have depinned and undergo quasi-free motion.
In this regime, interactions with the pinning centers produce a velocity
oscillation as a function of time with a frequency
determined by the pinning density and applied current density.
For $K=-0.02J$
there is no skyrmion annihilation until $j\geq 4.7\times10^{10}\text{A/m}^2$,
above which all of the skyrmions annihilate.

We also investigated the effects of the anisotropy strength
$K$ of the pinning sites. 
%with stronger pinning centers relating to larger values of $|K|$.
As $|K|$ increases, the onset of soliton motion shifts to higher values of $j$
since larger applied current densities are required to depin skyrmions from
stronger pinning centers.
The velocity-current
shape does not remain the same for all
values of $K$.
When $|K|\leq0.04J$, soliton motion extends over a range of $j$ values
and the velocity-current curve is
very similar in form to
previous particle-based studies of soliton motion in skyrmion chains
\cite{Vizarim22}.
On the other hand, for $|K|\geq0.06J$
the average skyrmion velocity behavior changes significantly
and soliton motion occurs only over very restricted values of $j$. 
The change is a consequence of skyrmion annihilation,
which destroys the soliton motion.
The annihilation profile varies as a function of $|K|$,
and the onset of complete skyrmion annihilation
shifts to higher values of $j$ with increasing $|K|$.
The annihilation process produces a reentrant pinning phase,
where skyrmions cease moving when the soliton is destroyed.
The reentrant pinning phase
falls between the soliton phase and the
quasi-free skyrmion motion phase.
%indicating that this reentrant pinning phase is a transient phase between the soliton and the quasi-free phases.
%CIJOL WORKING
Our results indicate that stable soliton motion in skyrmions is possible
and can occur under conditions where the effects of the skyrmion Hall angle
are strongly reduced. 
This could be useful
for technological applications by
providing precise and controllable information transmission
via soliton motion.

\subsection*{Credit authorship contribution statement}
{\bf J. C. Bellizotti Souza}: Investigation, Software, Images, Writing - Original Draft
{\bf Nicolas Vizarim}: Methodology, Investigation, Images,
Visualization, Writing - Review \& Editing.
{\bf Charles Reichhardt}: Conceptualization,
Methodology, Writing - Review \& Editing.
{\bf Pablo Venegas}: Supervision, Funding acquisition.
{\bf Cynthia Reichhardt}: Software, Methodology, Writing - Review \& Editing.

\subsection*{Declaration of Competing Interest}
The authors declare that they have no known competing financial interests
or personal relationships that could have appeared to influence the work
reported in this paper.

\subsection*{Acknowledgment}
This work was supported by the US Department of Energy through the Los Alamos National Laboratory. Los
Alamos National Laboratory is operated by Triad National Security, LLC, for the National Nuclear Security
Administration of the U. S. Department of Energy (Contract No. 892333218NCA000001). 
\\
J.C.B.S acknowledges funding from Fundação de Amparo à Pesquisa do Estado de São Paulo - FAPESP (Grant 2022/14053-8)
and Coordenação de Aperfeiçoamento de Pessoal de Nível Superior - CAPES.
\\
N.P.V acknowledges funding from Fundação de Amparo à Pesquisa do Estado
de São Paulo - FAPESP (2017/20976-3).

%\section*{References}
%\bibliographystyle{iopart-num}
\bibliography{mybib}

\end{document}